# Raman spectroscopy of optical phonon and incommensurate charge density wave modes in 2H-TaSe$_2$ exfoliated flakes

Lin Cui,[1,2] Qiaoling Huang,[1] Gaomin Li,[1] Yumeng You,[2] and Mingyuan Huang[1,*]

[1] *Department of Physics, Southern University of Science and Technology, Shenzhen 518055, P. R. China*

[2] *Ordered Matter Science Research Center, Southeast University, Nanjing 211189, P. R. China*

E-mail: huangmy@sustc.edu.cn

**Abstract:** 2H-TaSe$_2$ is a model transition metal dichalcogenide material that develops charge density waves (CDWs). Here we present variable-temperature Raman spectroscopy study on both incommensurate charge density waves (ICDW) and optical phonon modes of 2H-TaSe$_2$ thin layers exfoliated onto SiO$_2$ substrate. Raman scattering intensities of all modes reach a maximum when the sample thickness is about 11 nm. This phenomenon can be explained by optical interference effect between the sample and the substrate. The E$_{2g}^{ICDW}$ amplitude modes experience redshift as temperature increases. We extract ICDW transition temperature (T$_{ICDW}$) from temperature dependence of the frequency of E$_{2g}^{ICDW}$ mode. We find that T$_{ICDW}$ increases in thinner flakes, which could be due to a result of significantly enhanced electron–phonon interactions. Our results open up a new window for search and control of CDW of two-dimensional matter.

2H-Tantalum diselenide ($TaSe_2$) is one of the most extensively studied transition metal dichalcogenide (TMD) materials that exhibit charge density wave (CDW) transitions at low temperatures.[1-5] CDW is a periodic modulation of conduction electron densities and is usually found in metallic TMDs. 2H-$TaSe_2$ in bulk has transition from normal (metallic) phase to the incommensurate charge-density-wave (ICDW) phase at 122 K, followed by the commensurate charge-density-wave (CCDW) phase transition at 90 K.[6-9] However, the question of whether the ICDW of 2H-$TaSe_2$ is enhanced or suppressed upon thinning the samples down to few layers thickness has not yet been resolved.

Raman spectroscopy is a convenient and noninvasive technique in probing optical phonons and CDW phase transitions in various TMD materials.[10-14] Therefore, in this Letter we report our studies of $E_{2g}$ and $A_{1g}$ optical phonon modes and the ICDW mode ($E_{2g}^{ICDW}$) using variable-temperature Raman spectroscopy. We find that all Raman modes exhibit the strongest intensities when the sample thickness is ~11 nm. We explain this observation using optical interference effect. The ICDW modes redshift with increasing temperature. From temperature dependence of the frequency of the $E_{2g}^{ICDW}$ mode, we extract $T_{ICDW}$ in samples with different thicknesses. The increasing trend of $T_{ICDW}$ in thinner flakes, which could be due to a result of significantly enhanced electron-phonon interactions. Our results open up a new window for search and control of CDW of two-dimensional matter.

TaSe$_2$ single crystals used in this work were purchased from 2D Semiconductors Company. Ultrathin layers were exfoliated onto Si substrates with 80 nm SiO$_2$ by the standard "Scotch-tape" method.[15] Thicknesses of exfoliated TaSe$_2$ ultrathin layers were determined by atomic force microscopy (AFM). A representative AFM image and height profile are shown in Fig. 1b. Raman scattering was conducted on freshly cleaved samples that were mounted in a cryostat with a window for optical access. A Helium-Neon laser at 632.8 nm was used, and the laser power was kept below 0.2 mW to avoid heating effect. Two ultra-narrow band notch filters were used to suppress the Rayleigh scattered light. The scattered light was dispersed by a Horiba iHR550 spectrometer and detected by a liquid nitrogen cooled CCD detector. Temperature of the TaSe$_2$ samples was estimated by the ratio of Stokes and anti-Stokes Raman scattering intensities.[14, 16, 17] Variable temperature Raman spectra for each sample were taken during the warming process from ~10 to ~300 K.

Fig. 1a displays a representative optical image of an exfoliated flake. Areas with different thicknesses show color contrast. The unit cell of the normal phase ($D_{4h}^{6h}$) in 2H-TaSe$_2$ consists of two trigonal prisms of Se atoms including a Ta atom as shown in Fig. 1c. The unit cel contains two molecular units and the lattice vibration is reduced to the following normal modes:[8]

$$A_{1g}+ 2B_{2g}+E_{1g}+2E_{2g}+ A_{2u} +B_{1u}++2E_{1u} +E_{2u} \qquad (1)$$

The $A_{1g}$, $E_{1g}$, and $E_{2g}$ modes are Raman active.[3] In the low-temperature CCDW phase, a superlattice of $3a_0 \times 3a_0 \times c_0$ is formed.[18] The commensurate superlattice shown schematically in Fig. 1d is found in 2H-TaSe$_2$ below low-temperature CCDW phase. The red open circles are Bragg points of the high-temperature structure; red solid circles are the primary superlattice Bragg points of the commensurate superlattice. The structure has the same space group ($D_{4h}^{6}$) as the original lattice and the unit cell contains 18 molecular units. Six points on the $\Sigma$ lines and the two k points in the original phase are folded into the $\Gamma$ point of the commensurate phase. Moncton revealed the Kohn anomaly of the $\Sigma_1$-symmetric LA-phonon branch on the $\Sigma$ line by neutron scattering.[18, 19] In 2H structure, the LA branch is degenerate with the $\Sigma_1$ rigid-layer mode in the largest part of the $\Sigma$ line. The 12 $\Sigma_1$ modes reduced to $2A_{1g} + 2E_{2g} + 2B_{1u} + 2E_{2u}$ modes in the CCDW phase and the two $K_6(E_2)$ modes, where Ta atoms are displaced in the basal plane, reduced to $2E_{2g} + 2E_{1u}$.

Fig. 2a shows Raman spectra of TaSe$_2$ flakes with different thicknesses at around 10 K, which is well below the $T_{CCDW}$. These samples were exfoliated on the same Si wafer and measured under the same condition. To explore intrinsic CDW properties of real two-dimensional 2H-TaSe$_2$, high quality samples with thickness reaching the monolayer limit are necessary. The 1-nm-thick sample is the thinnest we have obtained so far. However, In the case of the 2-nm-thick sample, no Raman peaks are observed in below 100cm$^{-1}$. A difference in the background intensity of TaSe$_2$ flakes with different thicknesses can however be noticed, which could be attributed to fluorescence (see in the SI).

Fig. 2b shows polarized Raman spectra from a 32-nm-thick flake at around 10 K. The notation Z(XX)Z indicates the wave-vector and polarization direction of the incident light, and polarization and wave-vector direction of the scattered light from the left side. Both the A$_{1g}$ and the E$_{2g}$ modes are observed in the Z(XX)Z configuration, A$_{1g}$ modes are observed in the Z(YY)Z configuration, and E$_{2g}$ modes are observed in the Z(YX)Z configuration. The phonon energies of five modes below 100 cm$^{-1}$, the E$_{2g}^{ICDW}$ mode at 54 cm$^{-1}$ and A$_{1g}^{CCDW}$ mode at 77 cm$^{-1}$ at low temperatures show large softening toward the CDW to the normal-phase-transition temperature (see Fig. 3a).[1,3] The E$_{2g}$ mode of 64 cm$^{-1}$ at low ternperatures becomes very weak. These three modes come from the Kohn-anomaly modes on the Σ lines in the original phase. The two E$_{2g}$ modes at 87 and 79 cm$^{-1}$ are assigned to come from the K points in the original phase.[2] Moreover, the A$_{1g}^{CCDW}$ mode at 77 cm$^{-1}$ and E$_{2g}$ mode at 79 cm$^{-1}$ merge at around 10 K. The sharp peaks at 221 and 241 cm$^{-1}$ are present in the spectrum even at room temperature (see Fig. 3a) and are assigned to the E$_{2g}$ and A$_{1g}$ optical phonon modes, respectively.[15]

Fig. 2c shows the area intensities of E$_g$ and A$_{1g}$ Raman modes (shown in Fig. 2a) as a function of thickness. With decreasing thickness Raman intensities of all prominent peaks experience a significant increase first and then followed by a decrease. Intensities of the ICDW and CCDW Raman peaks show the same trend. The Raman intensity reaches a maximum when the sample thickness is ~11 nm whose intensity is about 4 times as strong as that from the thick flakes. This nonmonotonic dependence of Raman intensities on the sample thickness is understood to be caused by the optical interference between TaSe$_2$ flakes and SiO$_2$/Si substrate, similar to that seen in graphene and MoS$_2$ atomic layers.[20, 21]

To quantify the observed nonmonotonic dependence of Raman intensities on the thickness of TaSe$_2$ layers, we use a multireflection model (MRM) of the excitation and scattered light. In this model, light undergoes an infinite number of reflections and refractions at the boundaries of TaSe$_2$ and SiO$_2$ layers (see the inset of Fig. 2c). A strategy for solving this problem is to calculate the effective reflection coefficient at the TaSe$_2$/SiO$_2$ interface by accounting for multiple reflections in the SiO$_2$ layer, and then analyze the light distribution in the TaSe$_2$ specimen layer.[21] The choice of the thickness of SiO$_2$ layer is usually close to 1/4 or 3/4 of the wavelength of the excitation light, which is the anti-Fabry-Perot resonance condition, to increase the effective reflectivity at the TaSe$_2$/SiO$_2$ interface. After the electric field of the excitation light in the TaSe$_2$ layers is calculated, the scattered light that can reach its surface is evaluated by the same method. The output Raman intensity from the TaSe$_2$ layer can be expressed as:[22]

$$I = \int_0^{d1} \left| F_{ex}(x) F_{sc}(x) \right|^2 dx, \qquad (2)$$

in which $x$ is the depth that light travels in the sample and it varies from 0 to d1 which is the thickness of the flake, $F_{ex}(x)$ and $F_{sc}(x)$ are the electric field amplitudes for the excitation light and the scattered light which can reach the surface, respectively. Detailed expressions for $F_{ex}(x)$ and $F_{sc}(x)$ are included in the SI. The calculated Raman scattering intensity is plotted as a function of thickness in Fig. 2c. The trend of intensity variation agrees with our experimental data, and the theoretically predicted maximum intensity occurs at around 10.5 nm, which agrees very well with our experimental value of ~11 nm.

On the other hand, The ratio of the integrated Raman intensities of the $A_{1g}$ mode to the $E_{2g}$ mode also shows distinctive behavior for different thickness samples (Fig. 2d). The origin of this difference is unclear, but it should not arise from the optical effects described above since they will affect both modes almost identically. Moreover, $E_{2g}$ mode exhibits redshift when increasing the thickness (Fig. 2d). An anomalous behavior of the $E_{2g}$ mode has been previously reported in few-layered $MoS_2$ and $WS_2$ films[23-26] and it might be caused by a stronger dielectric screening of the long-range Coulomb interactions between the effective charges in thicker samples.[27] A change in dielectric screening with thickness is also expected for $TaSe_2$.

Figure 3a shows Raman spectra from a 8-nm-thick flake at different temperatures. It is seen that with temperature increasing, $E_{2g}^{ICDW}$ mode show intensity weakening because the CDW lattice loses coherence as temperature approaching phase transition. Moreover, the $E_{2g}^{ICDW}$ amplitude modes experience redshift as temperature increases. We focus on the temperature range below 100 K to extract the frequencies of the $E_{2g}^{ICDW}$ since it is better defined in this temperature range. To quantify the ICDW transition temperature, peak positions of the $E_{2g}^{ICDW}$ mode of different TaSe$_2$ flakes were extracted from Lorentzian fitting of the data and plotted as a function of temperature in Fig. 3b. We choose the E$_{2g}^{ICDW}$ mode to characterize the transition temperature because this mode is well-defined and has a narrow linewidth. The phonon frequencies (peak positions) of the E$_{2g}^{ICDW}$ mode were fitted by the general power law expressed as:[10]

$$\omega(T) = \omega(0)(1 - T/T_{ICDW})^\beta, \tag{3}$$

where $\omega(0)$, the phonon frequency at 0 K, and $T_{ICDW}$ are fitting parameters; $\beta$ is a scaling parameter. According to the mean-field theory about ICDW mode softening, $\beta$ should be 0.5. However, many experiments show that $\beta$ has different values for different materials.[28, 29] Here we choose 0.175 for $\beta$ to match the $T_{ICDW}$=122 K of the bulk sample reported in the literature (details in the SI).[3] Extracted $T_{ICDW}$ values are plotted as a function of sample thickness in Fig. 3c. Clearly, $T_{ICDW}$ increases in thinner samples and reaches about 156 K for 4 nm flake, similar to that seen in 2H-NbSe$_2$,[11] which could be due to a result of significantly enhanced electron–phonon interactions.

**ACKNOWLEDGMENTS**

This work is supported by the Southern University of Science and Technology and the Shenzhen Municipal Committee of Science and Technology Innovation for fundamental research (Grant: No.: JCYJ20160531190446212).

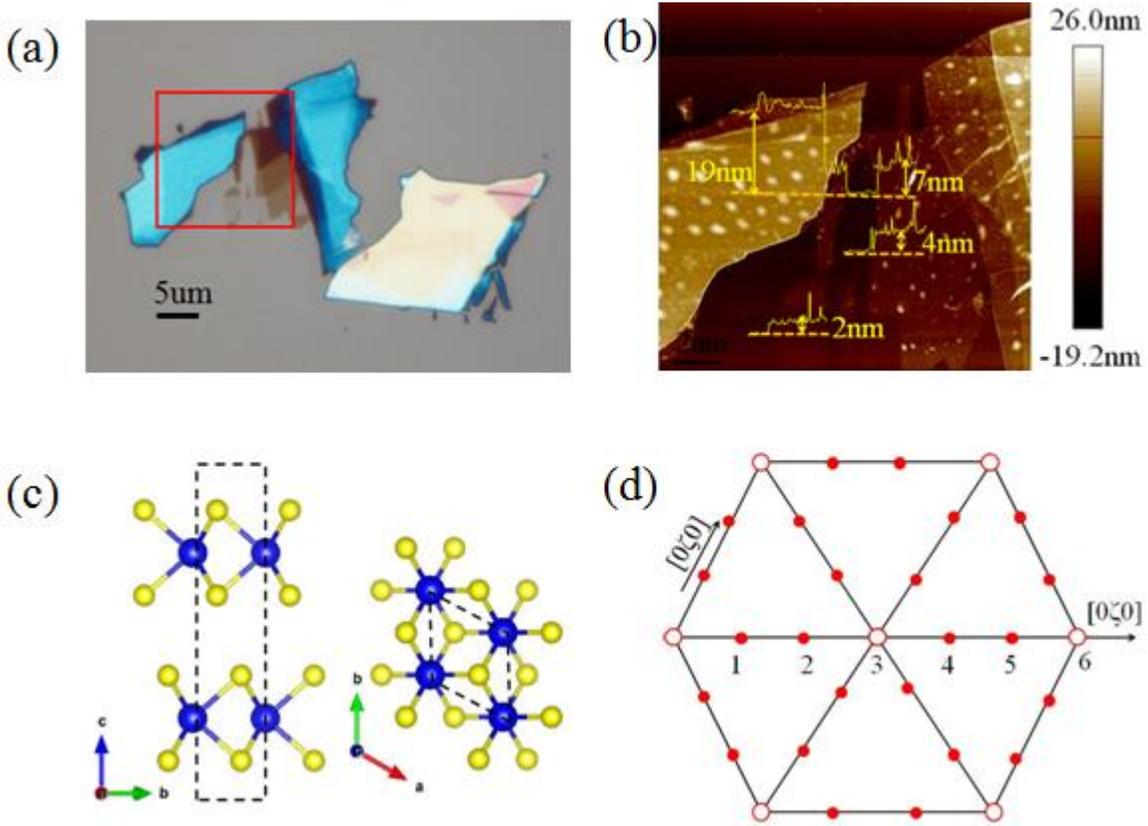

FIG. 1. (a) Optical image of an exfoliated flake. (b) AFM image of the boxed region highlighted in panel (a) and height profile along the vertical line. (c) Crystal structure of 2H-TaSe2. These views show the structures along the a-axis (at left) and c-axis (at right). (d) Reflections from the fundamental structure and from the superstructure in 2H-$TaSe_2$.

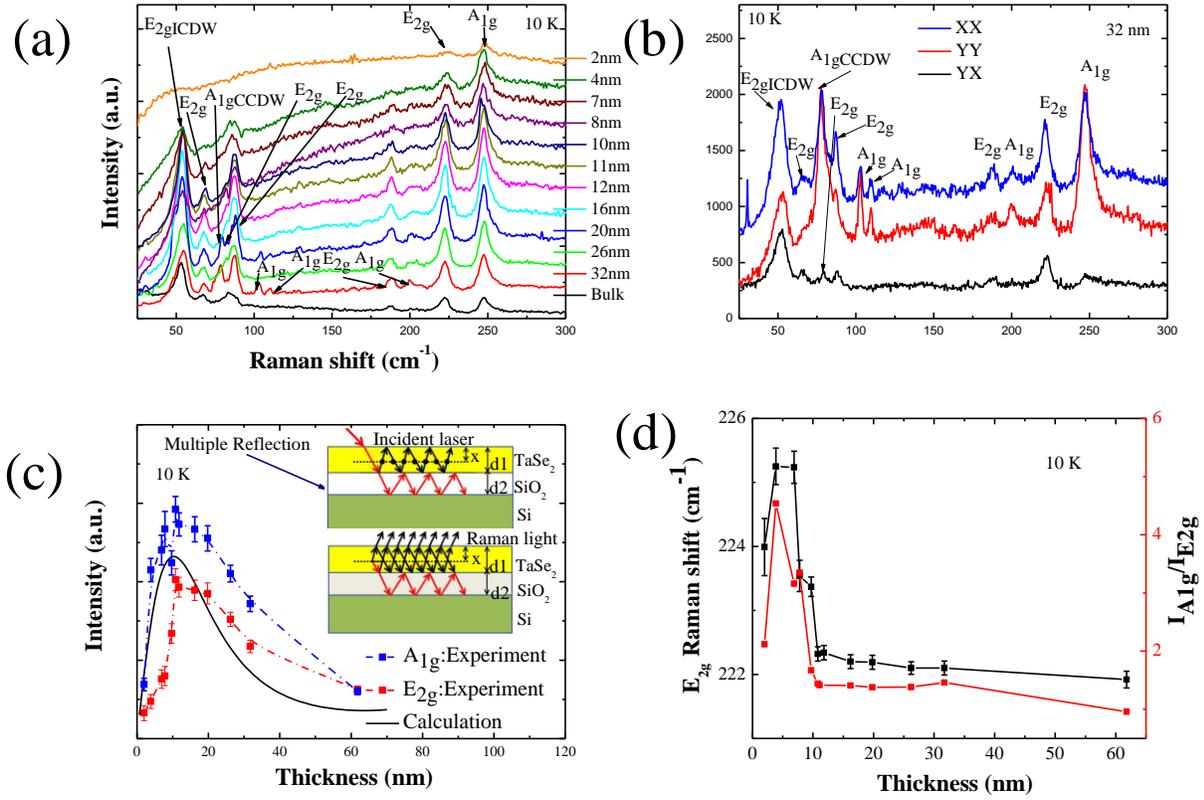

FIG. 2. (a) Raman spectra of TaSe$_2$ thin layers with different thicknesses at around 10 K. The spectra are shifted vertically for clarity. (b) polarized Raman spectra from a 32-nm-thick flake at around 10 K. (c) Thickness dependence of intensities of the E$_{2g}$ (red square) and A$_{1g}$ (blue square) optical phonon modes shown in panel (a). The dash lines are guides to the eye. The calculated overall intensity of the spectrum is plotted as the black solid line. The inset shows schematic diagrams for the optical paths of the excitation and scattered light. (d) Thickness dependence of the ratio of the integrated Raman intensities of the A$_{1g}$ mode to the E$_{2g}$ mode (red solid line) and the frequency of E$_{2g}$ mode (black solid lines).

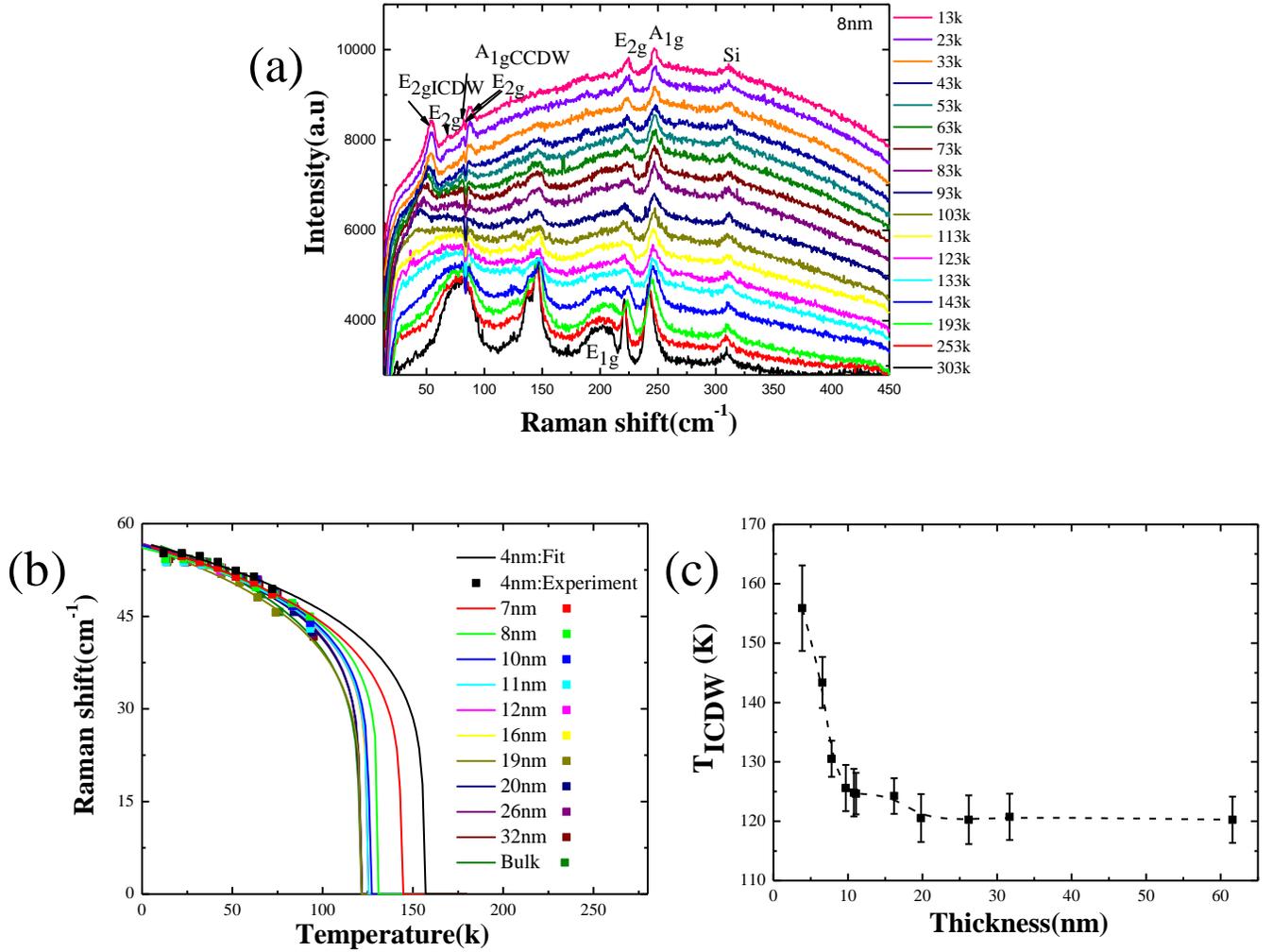

FIG. 3. (a) Temperature dependence of Raman spectra from a 8-nm-thick flake. Temperatures were determined by the ratio of Stokes and anti-Stokes Raman scattering intensities. (b) Temperature dependent peak positions of the $E_{2g}^{ICDW}$ mode. The solid lines are fittings by using the generalized power law discussed in the text. (c) Thickness dependence of the ICDW transition temperature of $TaSe_2$ flakes. The dash line is a guide to the eye.